\documentclass[conference]{IEEEtran}
\IEEEoverridecommandlockouts

\usepackage{cite}
\usepackage{amsmath,amsfonts}
\usepackage{graphicx}
\usepackage{endnotes}
\usepackage{enumitem,enumerate}
\usepackage{multirow} 
\usepackage{booktabs} 
\usepackage{algorithm} 
\usepackage{algpseudocode} 

\newcommand{\name}{STAR} 

\usepackage{soul}

\usepackage[a4paper, total={184mm,239mm}]{geometry}
\def\BibTeX{{\rm B\kern-.05em{\sc i\kern-.025em b}\kern-.08em
    T\kern-.1667em\lower.7ex\hbox{E}\kern-.125emX}}

\begin{document}

\title{STAR: An Efficient \underline{S}of\underline{t}max Engine for \underline{A}ttention Model with \underline{R}RAM Crossbar}

\author{\IEEEauthorblockN{
Yifeng Zhai$^{1}$, 
Bing Li$^{1}$,
Bonan Yan$^{2}$,
Jing Wang$^{3}$}
\IEEEauthorblockA{
\textit{Capital Normal University$^1$,}
\textit{Peking University$^2$,}
\textit{Renmin University of China$^3$,}\\
\textit{Email}: {\textit{zyifeng25@163.com,bing.li@cnu.edu.cn$^1$},
{\textit{bonanyan@pku.edu.cn$^2$}},
{\textit{jwang@ruc.edu.cn$^3$}}
}}
\thanks{Corresponding author: Bing Li, \textit{bing.li@cnu.edu.cn}
}}

\maketitle
\begin{abstract}
RRAM crossbars have been studied to construct in-memory accelerators for neural network applications due to their in-situ computing capability. 
However, prior RRAM-based accelerators show efficiency degradation when executing the popular attention models. 
We observed that the frequent softmax operations arise as the efficiency bottleneck and also are insensitive to computing precision. Thus, we propose \name, which boosts the computing efficiency with an efficient RRAM-based softmax engine and a fine-grained global pipeline for the attention models. Specifically, \name ~exploits the versatility and flexibility of RRAM crossbars to trade off the model accuracy and hardware efficiency.
The experimental results evaluated on several datasets show \name~achieves up to 30.63$\times$ and 1.31$\times$ computing efficiency improvements over the GPU and the state-of-the-art RRAM-based attention accelerators, respectively.  

\end{abstract}

\section{Introduction}
\label{sec:introduction}
Though some RRAM-based accelerators specialized for attention models have been discussed~\cite{Lu,ATT,ReTransformer}, they primarily focus on implementing the matrix multiplications on the RRAM crossbar. 
In this work, we observed the execution time of softmax operation grows quickly in attention models when the input sequence length increases. The latency of softmax exceeds that of matrix multiplication when the input sequence length is 512 in the BERT-base model, which reaches up to 59.20\% of the whole execution time.
Though our results are observed on a GPU platform, the softmax latency problem would be exacerbated on the RRAM-based accelerators because the matrix multiplication is significantly optimized by being implemented in RRAM crossbars~\cite{li2018reram} but the softmax still runs on the same circuits.
Thus, it is of significance to tailor an efficient softmax engine in RRAM-based accelerators for attention models.
To this end, we propose \name, which features an RRAM-based softmax engine by exploring the versatility and flexibility of RRAM crossbars to balance to the computing precision and efficiency. Moreover, an enhanced pipeline to balance the matrix multiplication and softmax operation in the attention is introduced.
The effectiveness of \name~is verified by the comparison results with the recent RRAM-based accelerators for attention models~\cite{ReTransformer}.

\section{RRAM-based Softmax Engine}
\label{sec:method}

\name~is primarily composed of two types of crossbar-based processing engines: \textit{MatMul~engine} for the VMM-dominated operations and \textit{Softmax~engine} for the softmax operation, respectively. 
The MatMul engine follows the design in ReTransformer~\cite{ReTransformer}.
As for the Softmax engine, different function units based on RRAM crossbars cooperate with each other to complete the softmax operation.
The Softmax engine has two distinct stages, $x_i-x_{max}$ and the exponential operation, which desire crossbars having different functions.

\subsubsection{$x_i-x_{max}$} The $x_i-x_{max}$ is achieved by one crossbar in a time-multiplex manner to complete the finding maximum and subtraction, respectively. Thus, the crossbar is denoted as CAM/SUB crossbar.
\begin{figure}[tb]
\small
    \centering
    \includegraphics[width=.8\linewidth]{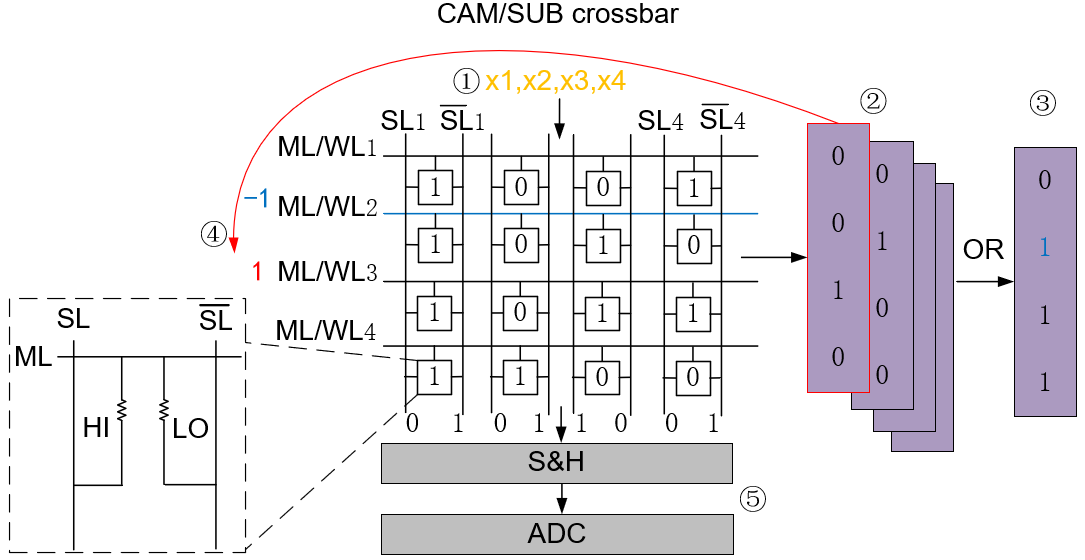}    \caption{The $x_i-x_{max}$ operation design.}
    \label{fig:findmax}
    \vspace{-12pt}
\end{figure}

Fig.~\ref{fig:findmax} shows the workflow of a 4$\times$8 CAM/SUB crossbar to find out the max value in $[x_1\cdots x_4]$.
The crossbar works as a CAM first. 
For each $x_i$, all the $WL$s of the crossbar are searched in parallel and the matchlines output a one-hot vector in which `1' denotes the matched line.
For example, if the data stored in the $WL_3$ in Fig.~\ref{fig:findmax} is consistent with $x_1$, the output vector would be [0,0,1,0](\textcircled{2}).
The outputs of matchlines cascade the $OR$ gates that merge the search results of all input $x_i$(\textcircled{3}).
Because the data are stored in descending order in the CAM crossbar, the index of the first `1' in the result vector corresponds to the row number of CAM storing the $x_{max}$. 
In the example of Fig.~\ref{fig:findmax}, $x_{max}$ stores at $WL_2$. 
Next, the crossbar executes the subtraction $x_i-x_{max}$. 
The match vector outputs 
will be used as the input voltage vector. Instead, the input for the $x_{max}$ row is a negative voltage (\textcircled{4}). Thus, the output from the $SL$s represents the results of $x_i-x_{max}$(\textcircled{5}).
\begin{figure}[tb]      
    \centering
    \includegraphics[width=.8\linewidth]{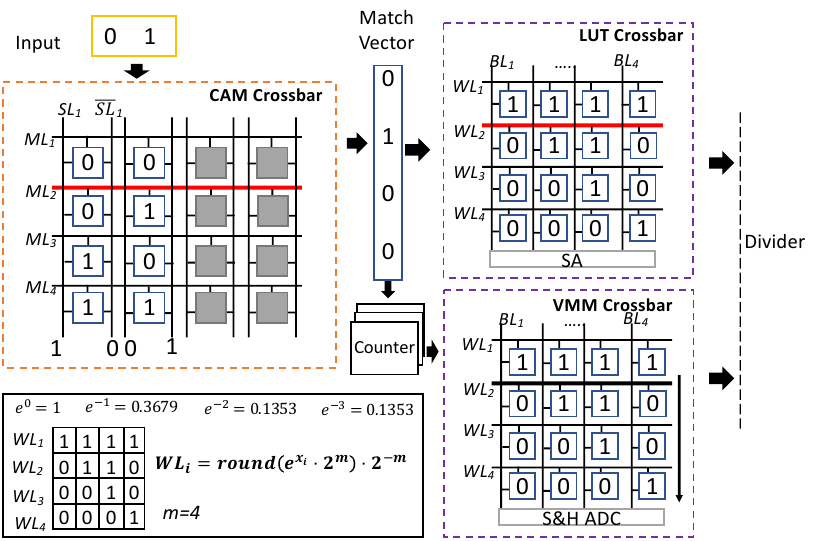}
    \vspace{-6pt}
    \caption{The exponential operation design in our softmax engine.}
    \label{fig:index}
    \vspace{-12pt}
\end{figure}
\subsubsection{Exponential Operation} The exponential operation is implemented by CAM crossbar and LUT crossbar. A VMM crossbar collaborates with them to complete the summation in the softmax.
All possible values of $x_i-x_{max}$ and their exponential results are preloaded in CAM crossbar and LUT crossbar, respectively.
Since the $x_i-x_{max}$ is always negative, we remove the sign bit to save the area of CAM crossbar.
Each input enters CAM crossbar and the output from the LUT crossbar is its exponential result.
At the same time, the match vector for CAM crossbar is sent to the counter for accumulation. When all $x_i$ complete the exponential computation, the results of the counter are sent to the VMM crossbar which stores exactly the same values as LUT crossbar to compute $\sum^{d}_{j=1}e^{x_j-x_{max}}$. 
Then the outputs of LUT crossbar and VMM crossbar enter the divider to complete the final division in the softmax.

Since the efficiency of the proposed softmax engine relates to the computing precision determined by the attention model, we analyzed the data range of all $x_i$ across three popular datasets for the BERT-base model such that balances the computing precision and hardware efficiency with \name. 
To achieve high model accuracy, the required bitwidth for CNEWS, MRPC, and CoLA are 8 bits (6-bit integer, 2-bit decimal), 9 bits (6-bit integer, 3-bit decimal), and 7 bits (5-bit integer, 2-bit decimal), respectively.

With the proposed RRAM-based Softmax engine, we introduce a vector-grained pipeline to improve the execution parallelism and efficiency for attention models. Thanks to the crossbar-based softmax engine, the complete attention mechanism operations could be in parallel in the vector granularity rather than the operand granularity in previous work.

%
\section{Experimental Results}
\label{sec:experiment}

\begin{table}[tb]
	\centering
	\small
	\caption{Comparison with the baseline CMOS-based softmax}
	\vspace{-6pt}
	\label{tab:softmax}
	\begin{tabular}{c c c}
		  \hline
		\textbf{Softmax Design} &\qquad\textbf{Area} &\qquad\textbf{Power}\\
		\hline
        Softermax~\cite{stevens2021softermax}&\qquad0.33$\times$ &\qquad0.12$\times$\\	
		\hline
		Ours (8-bit) &\qquad0.06$\times$ &\qquad0.05$\times$\\		
		\hline
	\end{tabular}
 	\vspace{-12pt} 
\end{table}


\begin{figure}[t]      
    \centering
    \includegraphics[width=.8\linewidth]{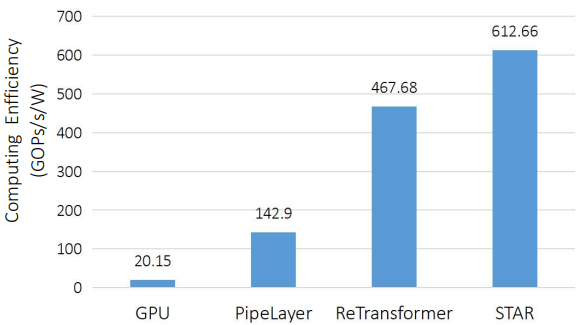}
    \caption{Computing efficiency comparison results.}
    \label{fig:STAR}
     \vspace{-0.3cm}
\end{figure}

We compared the proposed RRAM-based Softmax engine with an optimized COMS-based softmax accelerator, Softermax~\cite{stevens2021softermax} and a baselined CMOS-based softmax and compared \name~with a NVIDIA Titan RTX GPU platform and two ReRAM-based accelerators PipeLayer~\cite{song2017pipelayer} and ReTransformer~\cite{ReTransformer} to verify the collaboration of the proposed pipeline and Softmax engine.

The simulation of \name~is performed on NeuroSim~\cite{chen2017neurosim+} (for RRAM crossbar) and Synopsys Design Compiler (for the CMOS circuit), respectively. 
In the MatMul engine, the RRAM crossbar size is 128$\times$128 and the precision of ADC is 5-bit by referring to~\cite{ReTransformer}.
In the proposed Softmax engine, the size of the CAM/SUB crossbar is 512$\times$18 and the CAM (LUT, VMM) crossbar size is 256$\times$18 to support 9-bit data and computing precision.


Table~\ref{tab:softmax} is the comparison results of our Softmax engine with Softermax and the baseline CMOS-based softmax.  Here, the evaluated model is the BERT-base model on the CNEWS dataset with a sequence length of 128.
Compared to the baseline and Softermax, our Softmax engine is 0.06$\times$ and 0.20$\times$ smaller, respectively.
As for power, it achieves 0.05$\times$ and 0.44$\times$ power efficient than baseline and Softermax, respectively.
The results show our proposed Softmax engine offers a much better area efficiency and power efficiency than the baseline and Softermax.
Fig.~\ref{fig:STAR} compares the computing efficiency of GPU, Pipelayer~\cite{song2017pipelayer}, ReTransformer\cite{ReTransformer} and \name.
Computing efficiency here measures the number of operations that can be performed by a computing unit every unit time and every watt of power consumed.
\name~achieves the computing efficiency of 612.66GOPs/s/W.
Compared to GPU, Pipelayer and ReTransformer, \name~improves the computing efficiency by 30.63$\times$, 4.32$\times$ and 1.31$\times$, respectively.

\section*{Acknowledgement}
This paper is supported by the National Natural Science Foundation of China (NSFC) under grant No. 62204164, 62222411.

\bibliographystyle{ieeetr}
\bibliography{dac22}

\end{document}